# Machine-Learned Molecular Surface and Its Application to Implicit Solvent Simulations

Haixin Wei, Zekai Zhao, and Ray Luo

Departments of Materials Science and Engineering, Molecular Biology and Biochemistry, Chemical and Biomolecular Engineering, and Biomedical Engineering, Graduate Program in Chemical and Materials Physics, University of California, Irvine, California 92697, United States

## Abstract

Implicit solvent models, such as Poisson-Boltzmann models, play important roles in computational studies of biomolecules. A vital step in almost all implicit solvent models is to determine the solvent-solute interface, and the solvent excluded surface (SES) is the most widely used interface definition in these models. However, classical algorithms used for computing SES are geometry-based, thus neither suitable for parallel implementations nor convenient for obtaining surface derivatives. To address the limitations, we explored a machine learning strategy to obtain a level-set formulation for the SES. The training process was conducted in three steps, eventually leading to a model with over 95% agreement with the classical SES. Visualization of tested molecular surfaces shows that the machine-learned SES overlaps with the classical SES on almost all situations. We also implemented the machine-learned SES into the Amber/PBSA program to study its performance on reaction field energy calculation. The analysis shows that the two sets of reaction field energies are highly consistent with 1% deviation on average. Given its level-set formulation, we expect the machine-learned SES to be applied in molecular simulations that require either surface derivatives or high efficiency on parallel computing platforms.

## 1. Introduction

Electrostatic interactions play crucial roles in biophysical processes such as protein and RNA folding, enzyme catalysis, and molecular recognition. Thus, accurate and efficient treatment of electrostatics is vital to computational studies of biomolecular structures, dynamics, and functions. A closely related issue is the modeling of water molecules and their electrostatic interactions with biomolecules that must be considered for any realistic representation of biomolecules at physiological conditions. Implicit solvent model has been such an attempt, in which, the solute molecule is treated as a low dielectric constant region with a number of point charges located at atomic centers, and the solvent is treated as a high dielectric constant region. Among all the attempts, Poisson-Boltzmann equation (PBE) based implicit solvent models have proven to be among the most successful ones and are widely used in computational studies of biomolecules.

A crucial component of all implicit solvent models within the PBE framework is the dielectric model, *i.e.* the dielectric constant distribution of a given solution system. Typically, a solution system is divided into the low dielectric interior and the high dielectric exterior by a molecular surface. That is to say that the molecular surface is used as the dielectric interface between the two piece-wise dielectric constants. The solvent excluded surface (SES)[1-3] is the most used surface definition.[4, 5] Indeed, previous comparative analyses of PBE-based solvent models and TIP3P solvent models show that the SES definition is reasonable in calculation of reaction field energies and electrostatic potentials of mean force of hydrogen-bonded and salt-bridged dimers with respect to the TIP3P explicit solvent.[6-8] Given the complexity of the SES, one possible approach in adapting the SES in numerical solutions is to build the molecular surface analytically and then to map it onto a grid,[9-11] since analytical procedures can be time consuming. In fact, the analytical algorithms are mostly used in the visualization of SES.[12-22] Later, Rocchia *et al.* subsequently simplified the algorithm to facilitate the mapping of the SES to the grid.[5] This method is much faster, though it is less accurate and without an analytical expression.

The van der Waals (VDW) surface, or the hard sphere surface, represents the low-dielectric molecular interior as a union of atomic van der Waals spheres. This is a very efficient algorithm, though there exist many nonphysical high (solvent) dielectric pockets inside the solute interior when the VDW definition is used. Considering the limitation, the modified VDW definition was proposed. The basic idea of the modified VDW definition is to use the solvent accessible surface (SAS) definition for fully buried atoms and the VDW definition for fully exposed atoms.[23] However, the method is difficult to be optimized to reproduce the more physical SES definition.

The density approaches have recently been developed and can be used for numerical PBE solutions. Either a Gaussian-like function or a smoothed step function has been explored in previous developments.[24, 25] It has been shown that if the functional form is allowed to change, the density function can be explicitly optimized to reproduce the classical SES definition at least for certain "small" solvent probe radii, though this cannot be generalized to arbitrary probe radius values.[26] In this type of approaches, a distance-dependent density/volume exclusion function is used to define each atomic volume or the dielectric constant directly.[27-29] This is in contrast to the hard-sphere definition of atomic volume as in the VDW or the SES definition. Therefore, all surface cusps are removed by the use of smooth density functions.[24, 25]

In this study, we intend to address the difficulty of computing SES of complex molecular structures. The difficulties are due to the fact that SES is formed by different patches thus its analytical formulation must be piecewise and localized.[30] To overcome the difficulty, our solution is to use a machine learned analytical function. Indeed, in recent years, deep learning techniques have been widely adapted in handling multi-variable and highly non-linear functions similarly challenging such as SES. Deep learning/neural network is inspired by and resembles the human brain. The input variables are multiplied by the respective weights and then undergoes a transformation based on an activation function to obtain the outputs.[31] It has been mathematically proven that a single hidden layer was able to solve any continuous problem.[32] With all its advantages, deep learning has been applied in various biological fields, including gene expression[33-40], protein secondary and tertiary structures[41-53], protein-protein interactions[54-60], and others.[61]

In this study, we explored an SES method via the machine learning strategy, which is explicitly expressed as a level set function of atomic positions and radii. The level-set formulation is very convenient for many applications where surface curvatures or other higher-order surface parameters are needed. In addition, the new algebraic formulation is naturally suitable for parallel computing, like on the GPU platform, which would accelerate the implementation even further.

## 2. Method

2.1. Finite Difference Method for Solving PBE

As widely adopted for numerically solving partial differential equations, the finite difference method uses a uniform Cartesian grid to discretize the PBE. The grid points are numbered as $(i, j, k)$, where $i = 1, \ldots, x_m$, $j = 1, \ldots, y_m$, $k = 1, \ldots, z_m$, and $x_m$, $y_m$ and $z_m$ are the numbers of points along the three axes. The grid spacing between neighboring points can be uniformly set to *h*. To discretize the linearized PB equation,

$$\nabla \cdot \varepsilon \nabla \phi - \lambda \sum_i n_i q_i^2 \phi / kT = -4\pi\rho \qquad (1)$$

where the charge density $\rho$ can be expressed as $q(i,j,k)/h^3$, $q(i,j,k)$ is the total charge within the cubic volume centered at $(i, j, k)$, $\lambda$ is a masking function for the Stern layer. In the salt related term, $n_i$ is the number density of the ion of type *i* in the bulk solution, $q_i$ is the charge of the ion of type *i*, *k* is the Boltzmann constant, and *T* is the temperature. The final discretized PDE can be expressed as follows, if we ignore the salt related term as it is often modeled to be away from the molecular surface,

$$\begin{aligned}
\{&\varepsilon\left(i-\tfrac{1}{2},j,k\right)[\phi(i-1,j,k)-\phi(i,j,k)]\\
&+\varepsilon\left(i+\tfrac{1}{2},j,k\right)[\phi(i+1,j,k)-\phi(i,j,k)]\\
&+\varepsilon\left(i,j-\tfrac{1}{2},k\right)[\phi(i,j-1,k)-\phi(i,j,k)]\\
&+\varepsilon\left(i,j+\tfrac{1}{2},k\right)[\phi(i,j+1,k)-\phi(i,j,k)]\\
&+\varepsilon\left(i,j,k-\tfrac{1}{2}\right)[\phi(i,j,k-1)-\phi(i,j,k)]\\
&+\varepsilon\left(i,j,k+\tfrac{1}{2}\right)[\phi(i,j,k+1)-\phi(i,j,k)]\}/h^2 = -4\pi q(i,j,k)/h^3,
\end{aligned} \qquad (2)$$

where $\phi(i,j,k)$ is the potential at grid $(i,j,k)$, and $\varepsilon\left(i-\tfrac{1}{2},j,k\right)$ is the dielectric constant at the mid-point of the grids $(i,j,k)$ and $(i-1,j,k)$. All other $\phi$ and $\varepsilon$ are defined similarly here.

However, the simple discretization scheme above can be applied only to situations where the grid point and all its six neighbors are in the same region (regular grid points), either in solvent or in solute. Complication arises when at least one of the six neighboring grid points belongs to the different region (irregular grid points), i.e., at least one of the neighboring grid edges intersects the solute/solvent interface. Thus, how to define and determine the interface parameters is a key factor for setting up the linearized PBE. Without question, SES is the most adapted interface definition, as it was found to reproduce both energetic and dynamic properties of solvated molecules in explicit solvent.[6-8, 26] After choosing an interface definition, the discretization can then be handled with many different schemes. For example, the widely used harmonic average methods in a class of such strategies widely used in biomolecular simulations.[62-64]

2.2. Level Set Functions for Interface Definition

It is often more convenient to use a level set function to represent an interface. The level set function is often in the form of a second-order continuous distribution function $\varphi$, satisfying,

$$\begin{aligned}
\varphi(x,y,z) &< 0, when\ (x,y,z) \in \Omega^-\\
\varphi(x,y,z) &= 0, when\ (x,y,z) \in \Gamma\\
\varphi(x,y,z) &> 0, when\ (x,y,z) \in \Omega^+
\end{aligned} \qquad (3)$$

Here, $\Gamma$ represents the interface, $\Omega^-$ and $\Omega^+$ denote the inside and outside regions, respectively. For example, a signed distance function representing a sphere interface, centered at $(x_0, y_0, z_0)$ with a radius of $R$ can be expressed as,

$$\varphi(x,y,z) = \sqrt{(x-x_0)^2+(y-y_0)^2+(z-z_0)^2} - R. \qquad (4)$$

Not only level set functions can be used to determine the interface by setting up $\varphi = 0$, but also it can be used to compute various interface parameters, such as the normal and tangential directions on the interface,

$$\begin{aligned} \boldsymbol{\xi} &= (\varphi_x, \varphi_y, \varphi_z) \\ \boldsymbol{\eta} &= (\varphi_y, -\varphi_x, 0) \\ \boldsymbol{\tau} &= (\varphi_x \varphi_z, \varphi_y \varphi_z, -\varphi_x^2 - \varphi_y^2) \end{aligned} \tag{5}$$

Here, $\boldsymbol{\xi}$, $\boldsymbol{\eta}$ and $\boldsymbol{\tau}$ are the normal and two tangential directions, respectively. In addition, some advanced discretization schemes, like IIM,[65-68] also require higher-order interface parameters, such as surface curvatures that can also be computed as follows,

$$\begin{aligned} \xi_{\eta\eta} &= -\frac{\varphi_{\eta\eta}}{\varphi_\xi} \\ \xi_{\tau\tau} &= -\frac{\varphi_{\tau\tau}}{\varphi_\xi} \\ \xi_{\eta\tau} &= -\frac{\varphi_{\eta\tau}}{\varphi_\xi} \end{aligned} \tag{6}$$

In summary, level set functions are convenient mathematical tools in handling interface-related problems: they can be used to obtain the interface itself and all its geometry parameters in a straightforward manner. Thus, it would greatly benefit discretization of the PBE if we can express SES as a level set function. However, construction steps of SES are simply too complex to be reproduced algebraically. What we are trying to do here is to explore how to approximate SES with a level set function that is learned on modern computers.

2.3. Deep Learning and Neural Network

Artificial neural networks (ANNs), usually simply called neural networks, are mathematical models that have been motivated by the brain function.[31, 32] A simple example of an ANN structure is shown below in Figure 1.

The basic idea of a neuron model is that an input, $\boldsymbol{x}$, together with a bias, $b$, is weighted by $\boldsymbol{w}$, and then summed together[69], as shown schematically in Figure 1. The bias, $b$, is a scalar value whereas the input $\boldsymbol{x}$ and the weights $\boldsymbol{w}$ are vectors, i.e., $\boldsymbol{x} \in \mathbb{R}^n$ and $\boldsymbol{w} \in \mathbb{R}^n$ with $n \in \mathbb{N}$ corresponding to the dimension of the input. The sum of these terms, i.e. $z = \boldsymbol{w}^T\boldsymbol{x} + b$, forms the argument of an activation function, $f()$, resulting in the output of the neuron model,

$$y = f(z) = f(\boldsymbol{w}^T\boldsymbol{x} + b) \tag{7}$$

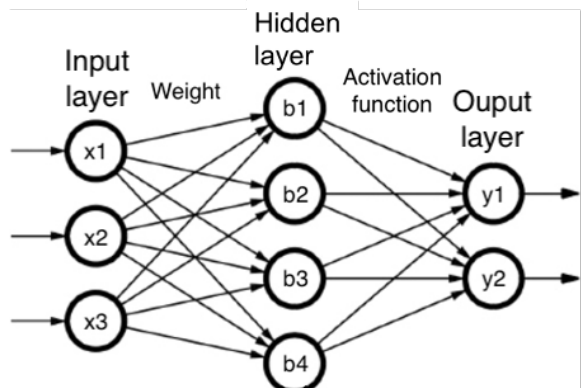

**Figure 1**. Structure of an artificial neural network with one hidden layer. Here, each circular node represents an artificial neuron, and an arrow represents a connection from the output of one artificial neuron to the input of another.

The role of the activation function, $f()$, is to perform a non-linear transformation of $z$. There are many activation functions in practice, such as sigmoid, Tanh, ReLU, Leaky ReLU, and so on, and the ReLU activation function is usually the most popular activation function for deep neural networks.[69]

Though its structure may seem highly complicated, a neural network can be viewed as a combination of simple elementary functions. Thus, a neural network is differentiable to any order, as all its component functions are differentiable to any order, except for a countable number of points. This is an important reason why we have chosen a machine learning approach to express the SES level set function, because it ensures that the SES level set function can be used to compute surface derivatives to any order as needed.

Due to its excellent mathematical properties, ANNs have shown great promises in a range of applications.[61] Most ANN applications fall into two categories: function approximation/regression analysis and classification problems. In this study, the ANN is applied as a classification tool because the goal is to know which region a grid point belongs to (solvent or solute region) in a PBE system.

### 3. Computational Details

The training data were generated from the Amber PBSA benchmark suite of 573 biomolecular structures.[70] A modified PBSA program was used to print out the training data. The default atomic cavity radii were read from the topology files. The solvent probe radius was set to 1.4 Å, the grid spacing to 0.95 Å, and all other parameters remain as default in the PBSA module in the Amber 20 package.[71] The training software package is Keras in TensorFlow, version 2.2.4.[72] Adam was chosen as the optimizer, and the squared hinge function as the loss function. The partition ratio of training and validating sets is 9:1, and an early stopping criterion of 100 epochs was used.

The training data contains two parts: labels or the target values, and the variables. The labels are integer values of irregular grid points derived from the Amber PBSA routine, which are calculated through a geometry-based SES algorithm.[63] The irregular points of the outside (solvent) region are labeled as -1, and those of the inside (solute) region are labeled as +1. The variables of the training data are the tuples of coordinates and the radii of those atoms near the irregular grid points, with the unit of Å. An atom is considered "near" a grid point if the distance between the grid point and the atom is within $R+2D_p$, where $R$ is the atom radius and $D_p$ is the diameter of the probe. The coordinates of the atoms are expressed in the relative frame whose origin is the grid point of interest. The nearby atoms were sorted based on their distances to the grid point before training. In the training data, the maximum nearby atoms are 48, so the maximum number of variables are 192. Overall, there are about 6 million entries of data generated from the Amber PBSA benchmark suite, which was separated into 6 subsets, each about 1 million entries.

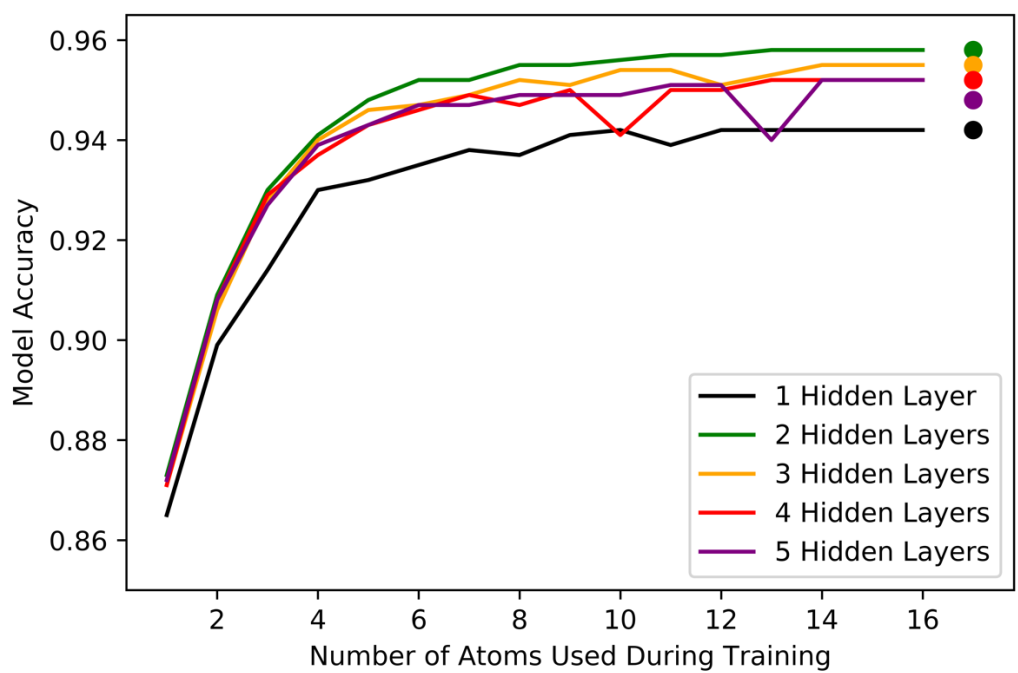

Figure 2. Incremental training of the model. The ANN model was set up with different numbers of hidden layers, each hidden layer is of 200 neurons. The model was trained with the first training data set. The analysis was also repeated with two additional training sets and the figures are shown in the **Supplemental Materials** Figure S1.

The training was conducted in three steps. First, the model was trained incrementally by adding one atom at a time, which means in the first round of training, the model was fed with only the first (nearest) atom's coordinates and radius. After the training reaches stabilization, the model was fed with the first two atoms' data, and so on. After enough number of atoms were fed to the atom, the accuracy of the model no longer improves. We then fed it with all the nearby atoms. In this step, we also investigated how many hidden layers are needed to reach stable prediction accuracy, starting with one layer to five layers. As shown in the Figure 2, the accuracy of the model becomes stable at 16 atoms, no matter how many hidden layers were used in the ANN. Therefore after 16 atoms, all available nearby atoms were fed to the model. Figure 2 further shows that 2 hidden layers are already very good for our problem.

The second step was to determine how many neurons are needed for the hidden layers. To determine the neurons needed in the first hidden layer, the number of neurons of both layers are varied, and the results are shown in Figure 3(a). Clearly, for the three tested training sets, all show that 100-neurons perform the best. Next, the first hidden layer was kept at 100 neurons, and the second hidden layer was varied, and the analysis is shown in Figure 3(b). Figure 3(b) shows that all tested conditions perform similarly, with 40 neurons slightly better than others so 40 neurons were selected.

The third step was to finalize the model. Given the model structure, which contains two hidden layers, with 100 neurons and 40 neurons respectively, we trained the model with all the training sets. To minimize GPU memory usage, the model was trained sequentially with one training set at a time. For the first training set, the model was trained incrementally by adding one atom at a time as discussed above. After the first training set, all nearby atoms were fed to the model in the subsequent training of remaining five sets. The training order among the training sets is irrelevant, because all training sets perform extremely similarly (for example Figures 2 and 3 and Figure S1). The training script and final model can be found at https:// http://rayluolab.org/ml-ses/.

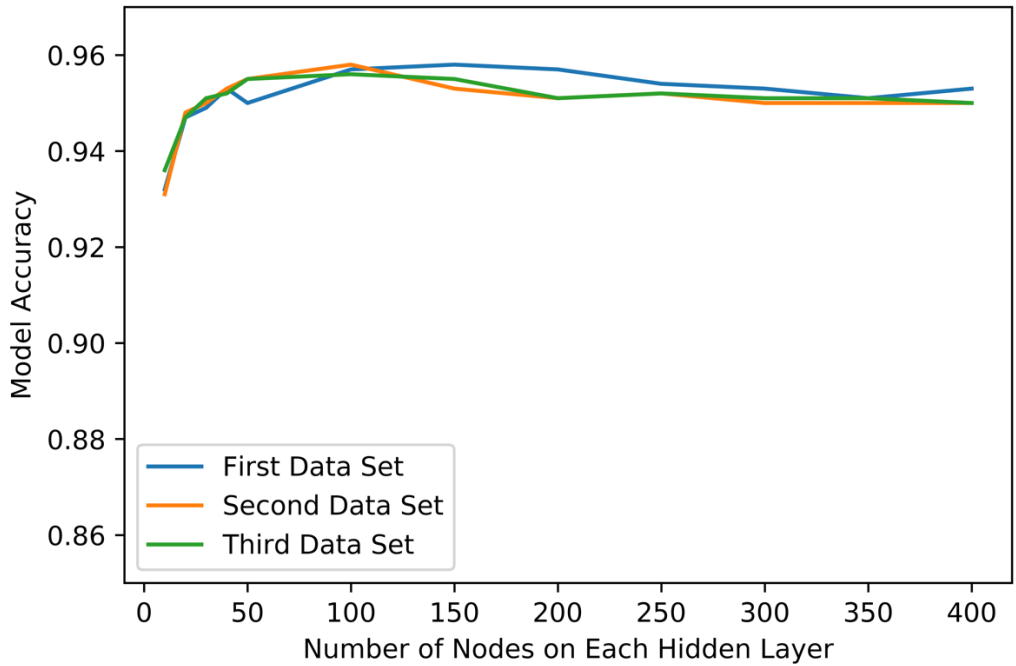

(a)

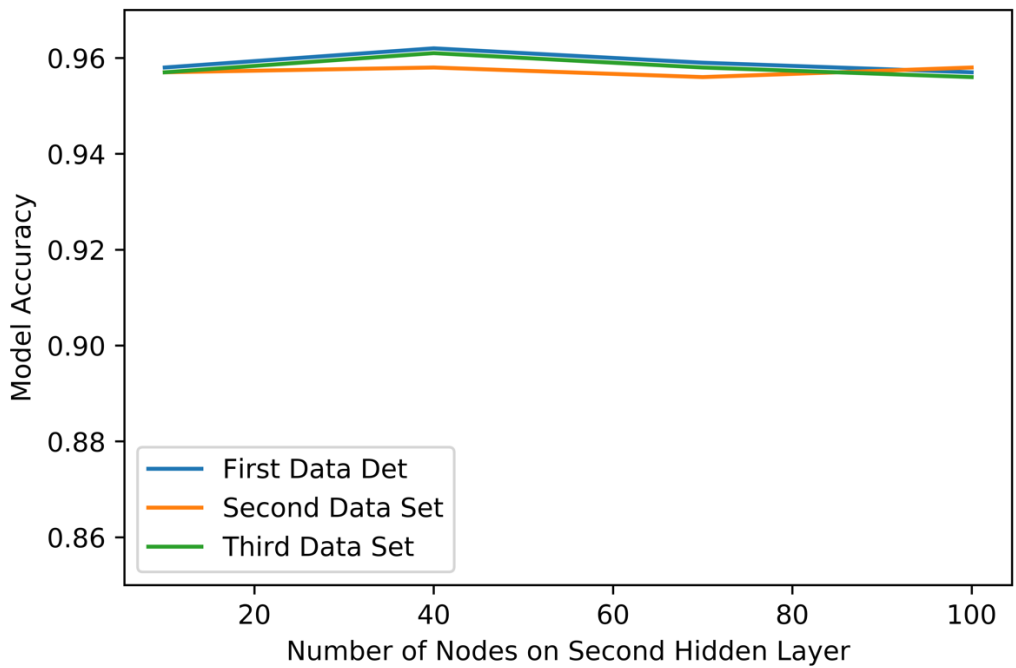

(b)

Figure 3. Determining of the number of neurons needed in the two hidden layers. (a) First layer analysis in the top panel. (b) Second layer analysis in the bottom panel.

## 4. Results and Discussion

4.1. Model Accuracy Analysis

To evaluate the overall performance of the machine-learned SES method, we first did an accuracy test on different structures. The test data sets were generated in the same fashion as the training data sets. Except for the original training protein monomers, we also included two extra sets of biomolecules for this analysis, nucleic acid structures[73] and protein/protein complex structures from previous PBSA developments.[74] The testing results are shown in Table 1.

| Data sets | Training data 1 | Training data 2 | Training data 3 | Training data 4 | Training data 5 | Training data 6 | Nucleic acid | Protein complex |
|---|---|---|---|---|---|---|---|---|
| Accuracy | 95.96% | 96.01% | 96.02% | 96.03% | 96.26% | 97.78% | 97.07% | 95.16% |

Table 1. Model accuracy with different structural data sets. The first six data sets are those from the training phase, and the last two sets are not included in the training phase.

Table 1 shows that the machine-learned SES performs very well for all structural sets, including those not included in training. The classification accuracies are basically around 96% for all data sets, with a variance about 1%. The method performs the best for training data set 6 and the nucleic acid data set. This is because those two contains mostly smallest molecules among all tested molecules. In general, a small molecule has simpler surface geometry so it is easier to predict their geometries. On the contrary, the protein/protein complex structures are much larger and thus have more complicated surfaces, so the method performs slightly worse on the data set, but still obtains an over 95% accuracy.

Another point worth mentioning is that the accuracy test clearly shows excellent transferability of the method, from the smaller training protein monomers to the nucleic acid structures and the larger protein/protein complex structures. The testing molecules overall do not resemble those in the training sets, such as the nucleic acids. Nevertheless, the method can successfully predict the SES of those unseen structures, showing that the method has indeed learned the fundamental rules for predicting the SES surface. In summary, the consistent accuracy confirms that the machine-learned SES method can be applied to typical computational analyses of biomolecules.

4.2. Reproducing Classical SES Molecular Surfaces

We chose six representative biomolecules and compared their molecular interfaces generated with three different methods: the newly developed machine-learned SES, the classical geometry-based SES, and a revised density function surface.[26] The superimposed surfaces of machine-learned SES and the classical SES are shown in Figure 4. The superimposed surfaces of machine-learned SES and the density function are shown in Figure S2 in the **Supplemental Materials**. Standalone surfaces from all three methods are also included in Figure S3-S5.

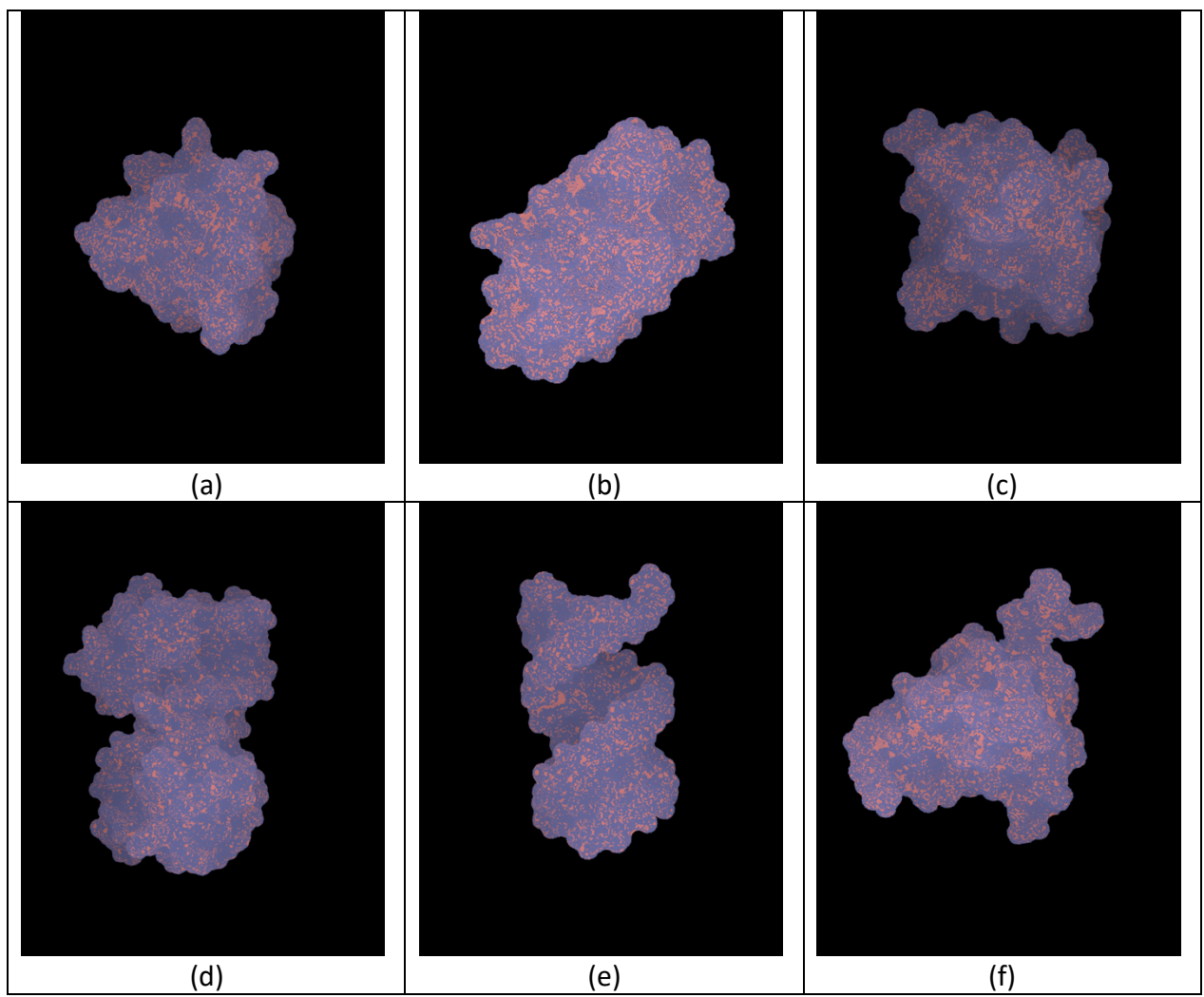

Figure 4. Superimposed rendering of machine-learned SES surface (blue) and the classical SES surface (red) of representative molecules. (a) PDB ID: 1enh, all-alpha protein; (b) PDB ID: 1pgb, all-beta protein; (c) PDB ID: 1shg, alpha/beta protein; (d) PDB ID: 1w0u, protein/protein complex; (e) PDB ID: 3czw, RNA duplex; (f) PDB ID: 3fdt, protein/DNA complex.

It is clear from Figure 4 that the machine-learned SES excellently reproduces the classical SES for all tested molecules, including both proteins and nucleic acids, consistent with the accuracy analysis shown in section 4.1. On the contrary, Figure S2 shows that the density function does not perform well in reproducing the classical SES for the tested systems. This is because the density function was found to agree well with the classical SES only with a smaller solvent probe, i.e., 0.7 Å. At the default solvent probe of 1.4 Å, the density surfaces are uniformly "fatter" than SES surfaces. Another limitation in the revised density function surface is that there are often deep reentry surface patches with very large curvature, which would cause numerical instability when it is used in numerical PB continuum solvents.[68]

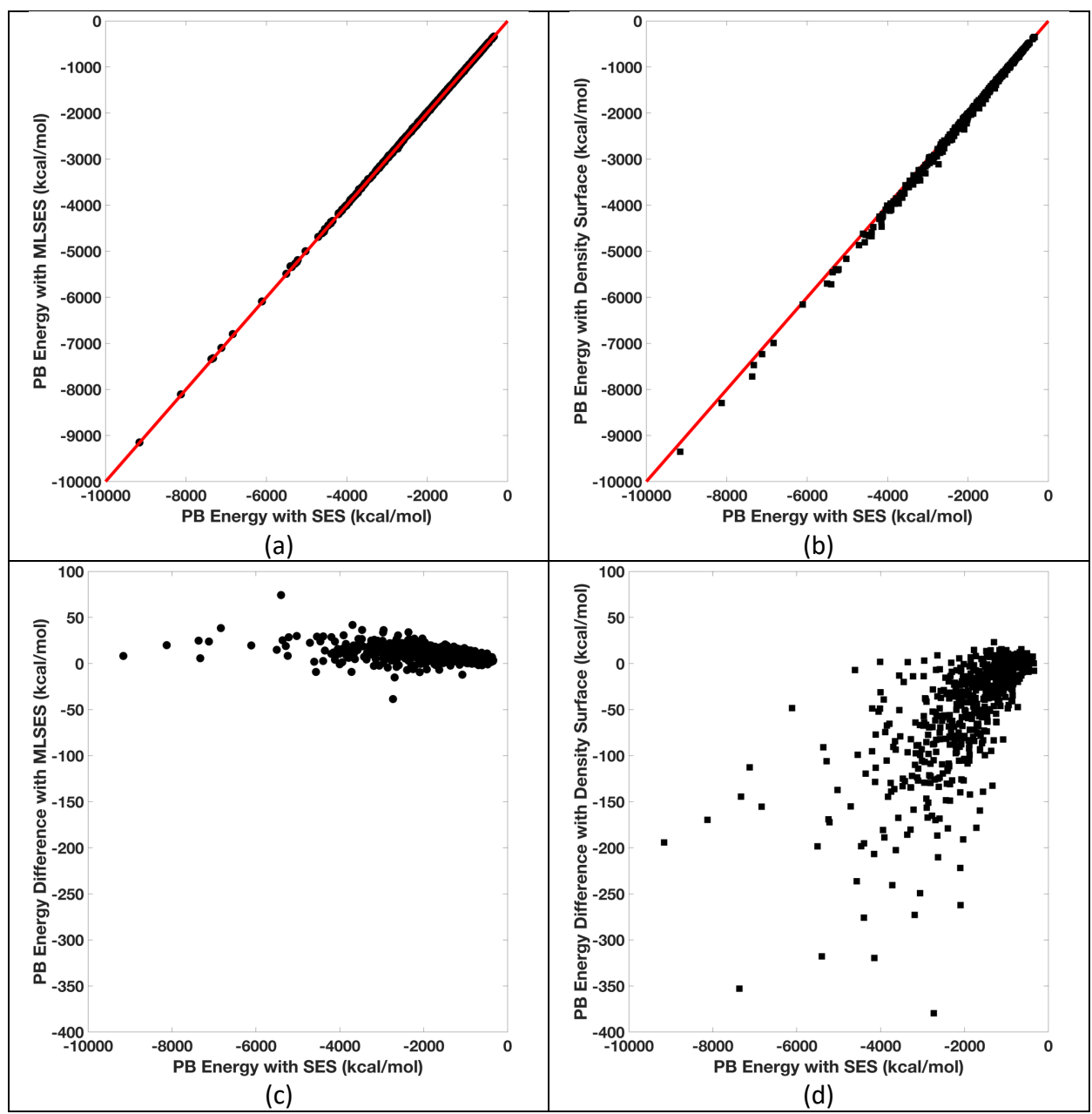

Figure 5. PB electrostatic solvation energies with different molecular surfaces. (a) Correlation between energies from machine-learned SES and classical SES. (b) Correlation between energies from density function and classical SES. (c) Energy differences in (a). (d) Energy differences in (b). The lines in (a) and (b) are the diagonal line "y=x" as reference.

4.3. Application to Poisson-Boltzmann Modeling

To illustrate the applications of the machine-learned SES for biomolecular studies, we implemented it into the AMBER/PBSA program. The PB reaction field energies were computed with the classical SES, the machine-learned SES, and the density function with otherwise identical conditions for all protein structures in the AMBER/PBSA benchmark suite.[70] Figure 5 shows the agreement between the computed energies with both machine-learned SES and the density function surfaces and those with the classical SES surface.

It is clear from Figure 5 that the machine-learned SES performs uniformly well when applied to the AMBER/PBSA program. The deviations between the PB energies with the machine-learned SES and those with the classical SES are around 1%. In contrast, the average deviation is roughly 2% but the maximum deviation is over 10% between the PB energies with the density function surface and those with the classical SES. Figure 5c/d further shows that the PB energies from the machine-learned SES are more random around the mean of zero but the PB energies from the density function are systematically more negative. This is because a small solvent probe (0.7 Å) has to be used in the density function to achieve reasonable agreement with the PB energies with the classical SES.[26] If the default probe of 1.4 Å is used, the PB energies would be 30% more positive than those with the classical SES.

4.4 Can machine learning extrapolate?

In this section, we want to discuss an interesting phenomenon observed during the training of the machine-learned SES. Since the ANN model was trained with irregular points near the solute-solvent interface (see section 3), it is obvious that the model performs well nearby the interface. However, it is unclear how the model performs away from the interface, for example at solvent-exposed grid points far away from the molecule or buried grid points deep inside the molecule interior, because we did not feed the model with such grid points and the model has never seen those situations.

Our initial impression was that the machine-learned SES model should also yield correct classification of the grid points in regions far away from the interface. The reasoning is simple as even an untrained student can classify these grid points without any calculation. If the model can successfully classify the most difficult irregular points, those far from the interface should be no problem. In another word, our machine-learned SES model should be able to extrapolate a little bit, or the model has obtained a certain level of "intelligence". The question is whether this is true.

To analyze the performance of the model in the entire solvation box, not just nearby the interface, we used the model to classify all grid points, include those grid points that are either far away outside the solute or deeply buried inside the solute. The performance of the classification is shown in Figure 6.

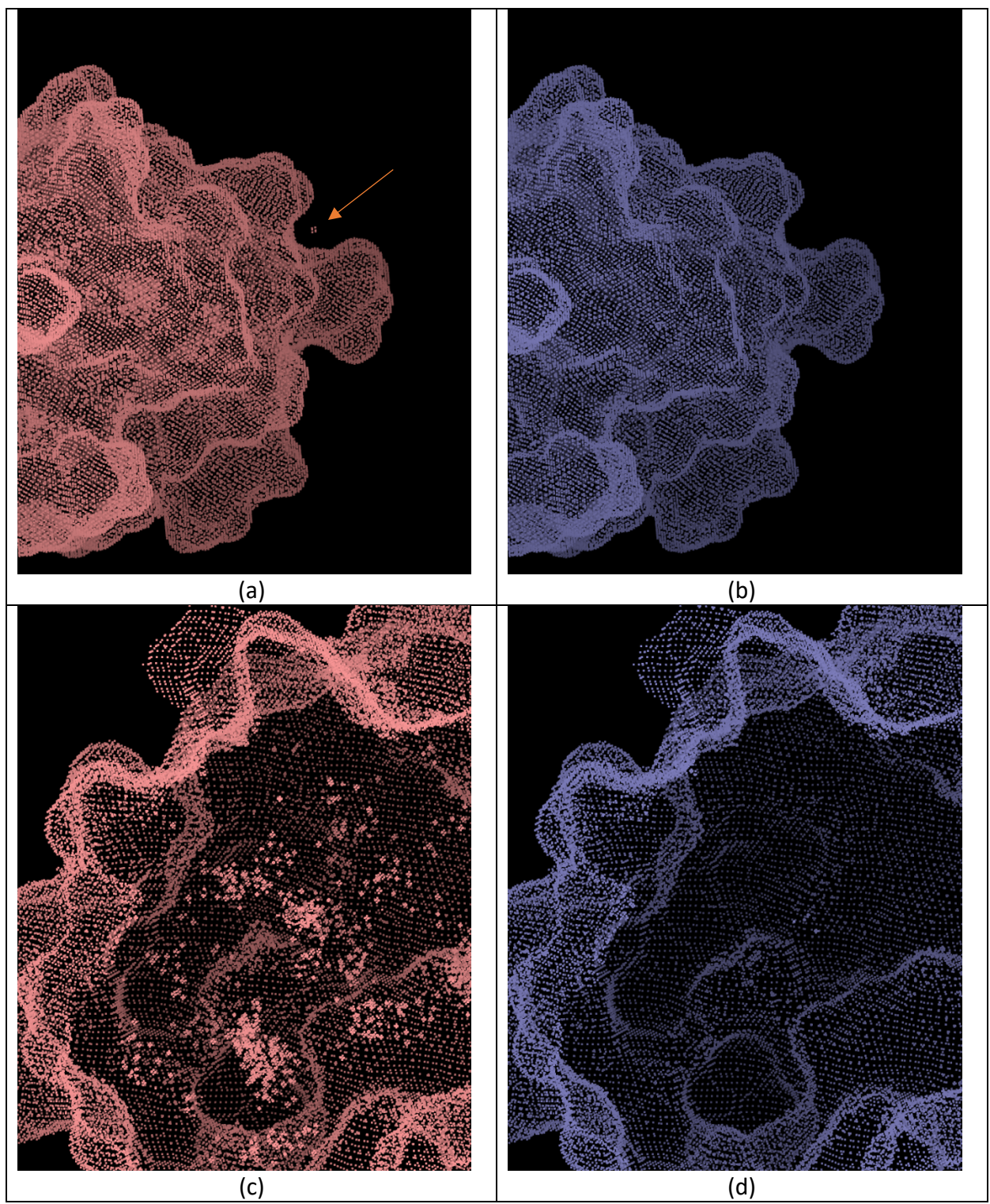

Figure 6. Detailed views of SES as predicted by first and second versions of the machine-learned SES model. The PDB id of the tested molecule is 1shg. (a) and (c) are from the first version of the model, outside vision and the inside vision, respectively. (b) and (d) are from a second version of the model, outside vision and the inside vision, respectively.

Figure 6 shows that the model clearly does not work at grid points far away from the interface. Specifically, Figure 6(a) illustrates the presence of an "island" or a small cluster of predicted interface grid points, outside the interface. Even more serious failures are within the molecule interior as illustrated in Figure 6(c), where hundreds of small clusters of interface grid points are wrongly predicted. This is surprising because the model can predict with an accuracy level over 95% in the most difficult interface region yet made hundreds of mistakes in the much easier regions.

These failures are due to the lack of similar data in the training of our initial model as only irregular grid points were fed to the training of the model. Thus, the remedy is to assembly a supplementary training set to include those grid points far from the interface (both inside and outside the molecule). Training of the initial model on the new data set leads to the second version of the model. Without surprise, the second version of the model is able to successfully classify those grid points incorrectly predicted by the initial model as shown in Figure 6(b)/6(d). Of course, it is never perfect and there are still few "islands" left as in Figure 6(d), but they can be ignored in most molecular modeling.

Our training experience of the SES model shows that, at least for the neural network we built, the machine-learned model does not possess any "intelligence". In other words, if we want the model work on grid points in a particular environment, we must train it in that exact environment. Basically, the model cannot extrapolate, even for the tested trivial situations. Thus, instead of artificial intelligence, "artificial memory" seems a more suitable terminology for the ANN model utilized here.

## 5. Conclusion

In this study, we developed a new level-set formulation to generate the solvent excluded surface through a machine learning process. SES is a widely used molecular surface definition and there are at least two advantages of expressing SES in a level set formulism: to facilitate parallel computing on GPU and to make it differentiable for future developments.

The level set was trained on the data generated from a set of heterogenous biomolecular structures, containing about 6 million entries. The training was conducted in three steps to determine the numbers of nearby atoms needed to define the level set, the number of hidden layers, and the number of neurons of each layer. After the final training, the machine-learned SES can predict the classical SES with an accuracy over 95%, on tested proteins, nucleic acids, and complex structures.

Analysis of visualized molecular surfaces of tested biomolecules shows that the machined-learned SES agrees excellently with the classical SES for the tested cases. It is also clear that a previous approach based on atomic density functions is clearly insufficient, generating a molecular surface often larger than the classical SES and causing deeper crevices in the reentry regions.

To further test the machine-learned SES, we implemented it into the AMBER/PBSA program to see if it can be used to reproduce the reaction field energies computed with the classical SES. Our results showed that the two sets of energies differ on average only about 1%, better than the 2% average deviation if the energies between the density-function surface and the classical SES are compared. Furthermore, the deviations in energies by the machine-learned SES are uniformly distributed yet the deviations in energies by the density-function surface often exhibit large deviations as high as 10%. This shows that the machine-learned SES is much more stable in reproducing the classical SES.

Finally, we highlight an interesting phenomenon in our training process: our testing shows that the model cannot classify grid points at locations that are not covered during the training process. To overcome the limitation, a supplementary training was conducted by feeding a new data set including the new locations. This finding shows that the ANN model cannot extrapolate. The observation should be kept in mind when future application of ANN models is used in structural analyses of biomolecules.

## 6. Acknowledgements

This work was supported by National Institute of Health/NIGMS (Grant Nos. GM093040 and GM130367)

## 7. Supporting Information

Supplemental materials include figures for incremental training of the ANN model for the second and the third data set; superimposed renderings of density function surfaces and classical SES of selected molecules/complexes; renderings of classical SES, machine-learned SES, and density function surfaces of selected molecules/complexes. Animation movies are also included for superimposed renderings of machine-learned SES and classical SES of selected molecules/complexes and for superimposed renderings of density function surface and classical SES of selected molecules/complexes. This information is available free of charge via the Internet at http://pubs.acs.org